\documentclass[reprint,
nofootinbib,
 amsmath,amssymb,
 aps,
]{revtex4-2}

\usepackage{graphicx}
\usepackage{dcolumn}
\usepackage{bm}

\usepackage[T1]{fontenc} 
\usepackage{booktabs} 
\usepackage{ dsfont }
\usepackage{float}

\usepackage{calrsfs}
\DeclareMathAlphabet{\pazocal}{OMS}{zplm}{m}{n}
\usepackage{xcolor}
\usepackage{pict2e}
\usepackage[percent]{overpic}
\usepackage{ upgreek }
\usepackage{xfrac}
\usepackage{abraces}
\usepackage{braket}
\usepackage{stmaryrd}
\usepackage{amsmath,empheq}
\usepackage{nicefrac}
\usepackage{abraces}
\usepackage{ctable}
\usepackage{upgreek} 
\begin{document} 

\preprint{APS/123-QED}

\title{Discovery prospects with the Dark-photons \\ \& Axion-Like particles Interferometer}
\author{Javier De Miguel$^{1,2,3}$}
 \email{javier.miguelhernandez@riken.jp}

\author{Juan F. Hernández-Cabrera$^{2,3}$}%

\author{Elvio Hernández-Suárez$^{2,3}$}%

\author{Enrique Joven-Álvarez$^{2,3}$}%


\author{Chiko Otani$^{1}$}%


\author{J. Alberto Rubiño-Martín$^{2,3}$}%

  \affiliation{
$^{1}$The Institute of Physical and Chemical Research (RIKEN), \\
Center for Advanced Photonics, 519-1399 Aramaki-Aoba, Aoba-ku, Sendai, Miyagi 980-0845, Japan}
\affiliation{$^{2}$Instituto de Astrof\'isica de Canarias, E-38200 La Laguna, Tenerife, Spain}
\affiliation{$^{3}$Departamento de Astrof\'isica, Universidad de La Laguna, E-38206 La Laguna, Tenerife, Spain}

\vspace{5pt}
\author{On behalf of the DALI Collaboration}
\date{\today}

\begin{abstract}
 We discuss the discovery potential of the Dark-photons \& Axion-Like particles Interferometer (DALI) in this letter. The apparatus, currently in a design and prototyping phase, will probe axion dark matter from the Teide Observatory, an environment protected from terrestrial microwave sources, reaching Dine--Fischler--Srednicki--Zhitnitsky-like axion sensitivity in the range 25--250 $\upmu$eV of mass. The experimental approach shows a potential to probe dark sector photons of  kinetic mixing strength in excess of several $10^{-16}$, and to establish new constraints to a stochastic gravitational wave background in its band. We identify different branches, including cosmology, stellar, and particle physics, where this next-generation halo-telescope may play a role in coming years.
\end{abstract}

\maketitle



\section{Introduction}
Non-luminous, `dark' matter is thought to play a role in galactic dynamics, the halos of spiral galaxies being benchmark astronomical laboratories \cite{ 1933AcHPh...6..110Z, 1970ApJ...159..379R, 2018RvMP...90d5002B}. In parallel, the current picture in cosmology suggests that a cosmological constant, $\Lambda$, and cold dark matter (CDM), enormously influence the evolution of the Universe. A highly accelerated inflationary expansion during a brief cosmological timeframe, and the cosmic microwave background (CMB), also shape the $\Lambda$CDM model \cite{1965ApJ...142..414D, 1965ApJ...142..419P, Starobinsky:1980te, Guth:1980zm, Linde:1981mu, 1982PhRvL..48.1220A, COBE:1992syq}. Large scale observations, simulations and the anisotropy measured in the CMB seem agreed in indicating the existence of CDM to favor the formation of the structures observed in the contemporary Cosmos \cite{blumenthal1984formation, davis1985evolution, 1993ApJ...416....1K, liddle1993cold, hernquist1996lyman, PhysRevD.80.023505, Hu_2001, spergel2000observational, doi:10.1146/annurev.astro.40.060401.093926, BOSS:2016wmc, Wang:2017wia, Libeskind:2017tun, 2020A&A...641A...6P}. 

The quantum chromodynamics (QCD) axion is a long-postulated  pseudo-scalar boson that arises as a consequence of the dynamic solution to the charge and parity (CP) symmetry problem in the strong interaction \cite{PhysRevLett.38.1440, PhysRevLett.40.223, PhysRevLett.40.279}. Furthermore, axion can simultaneously solve the dark matter enigma in a broad coupling strength to photons, ${g_{a\gamma\gamma}}$, to axion mass, $m_{a}$, parameter space \cite{ABBOTT1983133, DINE1983137, PRESKILL1983127,PhysRevD.98.030001}. Moreover,  beyond shaping galaxies by forming halos, the primordial density perturbations from which galaxies evolved may have been produced by the presence of temporary axion domain walls in the early Universe \cite{PhysRevLett.48.1156}. Indeed, approaches have been proposed that simultaneously solve the strong CP problem, axion dark matter, and inflation in an unique model \cite{Salvio:2015cja, ballesteros2017standard, ballesteros2017unifying, ernst2018axion, 2019FrASS...6...55R}. On the other hand, astronomical observations and simulations have given rise to the conjecture that dark matter in the nearby Universe is distributed in the form of substructures \cite{HOGAN1988228, moore1999dark, diemand2008clumps,2011MNRAS.413.1419V,2016JCAP...01..035T,2018MNRAS.475.1537M,vaquero2019early}. Lastly, a series of anomalous astronomical observations have led to hypothesize that new physics may play a role in stellar evolution \cite{2014PhRvL.113s1302A, Straniero:2015nvc, 2022JCAP...10..096D, 2022JCAP...02..035D}.

All of the above encourages the quest for axion dark matter. The search for dark matter by axion-photon interaction is promising. Laboratory experiments have excluded the sector $g_{a\gamma\gamma}\gtrsim10^{-7}$ GeV$^{-1}$ for $m_{a}\lesssim10^{-3}$ eV \cite{EHRET2010149, PhysRevD.88.075014, 2016EPJC...76...24D, PhysRevD.92.092002}. Helioscopes, partially overlapping stellar hints based on interactions of the axion to standard particles in the plasma of stars, exclude $g_{a\gamma\gamma}\gtrsim10^{-10}$ GeV$^{-1}$ for $10^{-2}\lesssim m_{a}/\mathrm{eV}\lesssim10^{5}$ \cite{CAST:2017uph}. Different astronomical campaigns and simulations rule out distinct parameter spaces \cite{PhysRevD.105.035022, PhysRevD.98.103015, Marsh_2017, PhysRevLett.118.011103, REGIS2021136075, 2021Natur.590..238B}; while cosmology also restricts the mass of this pseudo-Goldstone
boson to allow the structures observed in the contemporary Universe to be formed in a CDM picture and, as a result, the search for axion in the range 10$^{-6}\lesssim m_{a}/\mathrm{eV}\lesssim 10^{-3}$ is well motivated \cite{MARSH20161, PhysRevD.98.030001}.

The velocity dispersion of Halo dark matter is about $10^{-3}c$, with $c$ the speed of light. Therefore, the dynamic mass and the rest mass of these axions approximately coincide. The wavelength of the electromagnetic radiation emitted by axion-to-photon conversion is $\omega\sim m_{a}$. Axion haloscopes \cite{1983PhRvL..51.1415S} are magnetised detectors that probe Galactic dark matter via the inverse Primakoff effect \cite{Primakoff:1951iae}. This is the experimental approach to which we will confine our attention throughout this work; which is also sensitive to the dark photon via kinetic mixing \cite{Okun:1982xi, Vilenkin:1986}, and to high-frequency gravitational waves, within a faint parameter space determined by the Planck scale, through photon-graviton oscillation in an external magnetic field \cite{Chen:1994vk, Bastianelli:2007me, Dolgov:2012be, PhysRevLett.116.061102, 2019EPJC...79.1032E, 2020PhRvD.102j3501F}.

The rest of this paper is structured as follows. In Sec. \ref{II} we overview the experimental setup of the DALI Experiment. Section \ref{III} is devoted to examine its discovery potential. Conclusions are drawn in Sec. \ref{IV}.

\section{Experimental approach}\label{II}
The quest for axion at masses below a few dozen microelectronvolt has been performed by haloscopes for decades \cite{PhysRevLett.59.839, PhysRevD.42.1297, PhysRevLett.104.041301, PhysRevD.97.092001, PhysRevLett.128.241805, doi:10.1063/5.0098783}. Unfortunately, the search for axion at `high frequency,' above, say, two dozen microelectronvolt of mass or, equivalently, about half a dozen gigahertz, remains poorly explored. 
In the Sikivie cavity-haloscope principle, the signal power originating from axion-to-photon conversion scales as $P \propto g^2_{a\gamma\gamma}\, B_0^2 \, V C \, Q_L \, \rho_{a}/m_a$—e.g., \cite{PhysRevD.42.1297}. Here, $B_0$ is magnetic field strength; $V$ is the volume; $C$ is the form factor; $Q_L$ is the loaded quality factor and $\rho_a$ is the local density of axion dark matter, which saturates at $\sim$1/2 GeVcm$^{-3}$ \cite{2018RvMP...90d5002B}. The resonant frequency scales inversely on the cavity size, or the distance between the movable rods used for tuning, in the form $\nu_0\sim c/d$; $d$ being the diameter or distance. Smaller cavities provide a smaller detection volume that results in a lower signal power. Together with hurdles such as misalignment and sticking of the rods, radiation leakage as a result of external mechanical tuning, poor rod thermalisation, skin effect, etc., this can degrade the scanning speed at higher frequencies—e.g., see \cite{Stern:2015kzo,Quiskamp:2023ehr}. In contrast, DALI \cite{De_Miguel_2021, Hernandez-Cabrera:2023syh, Cabrera:2023qkt} incorporates a multilayer Fabry--Pérot resonator \cite{1899ApJ.....9...87P} instead of a closed resonant cavity, or rather than superconducting wire planes as proposed by Sikivie \cite{1983PhRvL..51.1415S, PhysRevD.50.4744, PhysRevD.91.011701}. In a dielectric Fabry-Pérot axion haloscope, the plate area, $A$, is decoupled from the resonant frequency, provided its size is larger than the scanning wavelength to avoid diffraction, enabling access to heavier axions. A similar concept is being developed by other collaborations, including MADMAX, ADMX-Orpheus, LAMPOST, MuDHI or DBAS \cite{MADMAXinterestGroup:2017koy, MADMAX:2019pub, PhysRevD.98.035006, PhysRevD.98.035006, Chiles:2021gxk, PhysRevD.106.102002, PhysRevLett.129.201301, PhysRevD.105.052010, PhysRevApplied.9.014028, PhysRevApplied.14.044051}. 

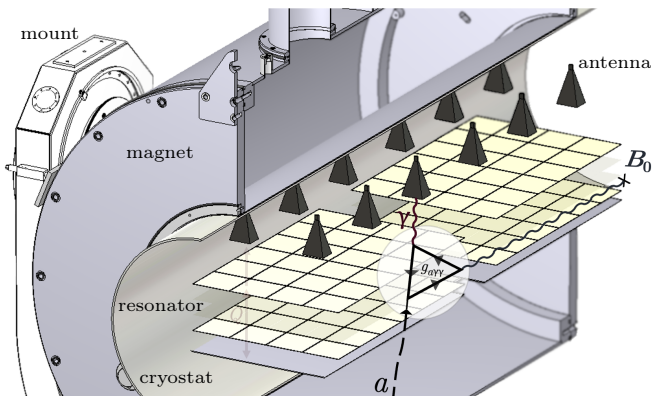
\begin{figure}[h]\centering
\includegraphics[trim=0 15 0 0,clip=true, width=.49\textwidth]{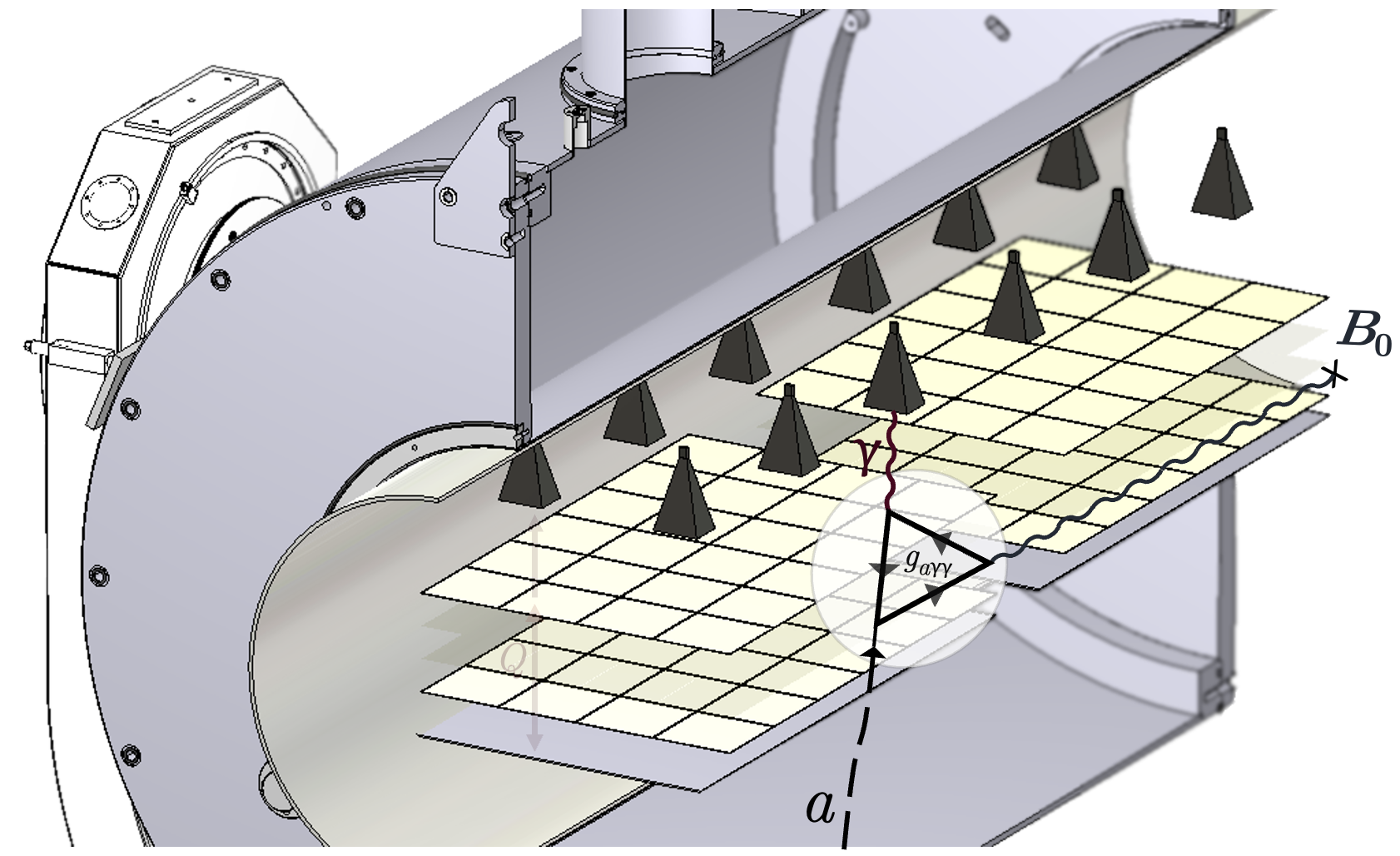}
\begin{scriptsize}
\put(-195,5){cryostat}
\put(-200,90){magnet}
\put(-203,32){resonator}
\put(-29,124){antenna}
\put(-240,135){mount}
\end{scriptsize}
\caption{DALI concept. The experiment cryostat, cylindrical shape, is housed within the warm bore of a solenoid superconducting magnet and employs an independent $^3$He cooling system which provides a sub-kelvin background temperature. The ceramic layers, in yellow, consist of a grid of wafers of a good dielectric material, e.g. zirconia (ZrO$_2$). Zirconia is a robust ceramic distributed in a range of thicknesses that has a high dielectric constant, up to $\varepsilon_r\gtrsim 40$, and a low loss tangent, of the order of  $\mathord{\mathrm{tan}}\,\delta\sim10^{-4}$ at $\pazocal{O}(10)$ GHz frequencies \cite{10.1063/1.353963,supp}. A polished mirror is attached at the bottom to envelop the Fabry--Pérot interferometer. The signal enhancement over a narrow bandwidth centered at a resonant frequency is the quality factor, $Q$. The microwave signal, originating at an axion-photon-photon vertex, is received by an antenna array, in black. This array of receivers is housed inside the same experiment cryostat with the field lines aligned with the high electron mobility transistors to cancel spurious effects \cite{doi:10.1063/1.366000,BRADLEY1999137,doi:10.1063/1.5107493}. The data acquired from different channels are post-processed and combined using techniques similar to radio interferometry that we have developed over decades of CMB observations—e.g., \cite{1994Natur.367..333H, 10.1046/j.1365-8711.2003.06338.x, 2011A&A...536A...1P, 2015MNRAS.452.4169G}. This tool allows for calibration, correction of phase mismatch and compensate for differences between pixels, etc., with a negligible error  \cite{Bryan:10, thesis_vignana}. This setup allows for a larger cross-sectional area inside a regular solenoid-type magnet. The apparatus rests on an altazimuth mount, in white on the left, to provide the instrument with additional directional sensitivity. The overall dimensions are a few meters in length and about one meter in diameter. Some components have been removed for simplicity.} 
\label{fig_1}
\end{figure}

In the DALI interferometer, constructive interference is caused by reflection off a top mirror, originating a standing wave. The resonant frequency is tuned by setting a plate distance of a fraction of the scanning wavelength, typically $\sim$$\lambda/2$ with a plate thickness of $\sim$$\lambda/(2\sqrt{\varepsilon_r})$ in a half-wavelength stack. However, the resonance is periodic over $\sim$$4\nu_0$ intervals \cite{renk2012basics, Ismail:16}. A one-eighth wavelength configuration allows the plate distance to be shortened by about four times to probe a desired frequency range. Therefore, a $\sim$$\lambda/8$ spacing is to be used at lower frequencies in order to save room inside the magnet bore, while the usual $\sim$$\lambda/2$ configuration can be employed at higher frequencies—cf. \cite{Hernandez-Cabrera:2023syh, Hernandez-Cabrera:2024SPIE, supp} for a proof of concept. Furthermore, multiple axion masses, about four wavelengths distant, could be probed simultaneously by alternating receivers centered at different frequencies on the focal plane of the haloscope without significantly decreasing its performance, in analogy with CMB experiments to which we have contributed—e.g., \cite{1994Natur.367..333H, 10.1046/j.1365-8711.2003.06338.x, 2011A&A...536A...1P, 2015MNRAS.452.4169G}. Once implemented, this multi-frequency approach would double the scanning speed of the haloscope with respect to the numbers that we will adopt throughout this article. The quality factor, $Q$, is the signal power enhancement over a narrow bandwidth centered at a resonant frequency. The group delay time, $\tau_g$, is the average lifetime of a photon within each volume enclosed between two adjacent reflectors, $\tau_g=-d\phi/d\omega$; $\phi$ being the phase. The quality factor of a Fabry--Pérot resonator is $Q=\omega \tau_g$; with $\omega$ the angular frequency \cite{renk2012basics}.\footnote{The power boost factor as defined by MADMAX and the quantum optics definition of the quality factor are equivalent in a transparent mode. } The $Q$ factor scales linearly with the number of layers in series \cite{Millar:2016cjp, Hernandez-Cabrera:2023syh}. A higher electric permittivity, $\varepsilon_r$, results in a higher $Q$ and a narrower full width at half maximum of the spectral feature caused by autocorrelation. The signal originating from axion-to-photon conversion is weak. The quality factor necessary to achieve QCD axion sensitivity, $Q\!\sim\!10^{4}$, is tenable over a bandwidth of several dozens of megahertz at frequencies of tens of gigahertz \cite{De_Miguel_2021, Hernandez-Cabrera:2023syh}. The error budget in plate spacing is set to a fraction of a few hundredths of the scanning wavelength by means of electromagnetic finite elements method simulations with adaptive mesh refinement \cite{costa2010introduction}. In consequence, the experimental range of DALI extends up to approximately 250 $\upmu$eV axion mass. Over broad bands, the experiment requires reconfiguration. This reconfiguration involves replacing receivers every few gigahertz, mechanical tuner components, and other accessories. Taking into account a scan speed of a few GHz per year, this reconfiguration is planned every several years as a part of a scientific program that would extend over more than a decade.
The DALI concept is shown in Fig. \ref{fig_1}, and detailed in depth in \cite{De_Miguel_2021}.

In DALI, the signal power induced by axion scales as $P \propto g^2_{a\gamma\gamma}\, B_0^2 \, A \, Q \, \rho_{a}/m_a^2$. A highly sensitive DALI instrument requires to maximize the cross-sectional area of detection, $A$, via a large magnet bore; the power enhancement factor, $Q$, and, crucially, the external field strength, $B_0$, by incorporating a potent superconducting magnet. Niobium–titanium, NbTi, superconductor does not provide magnetic field flux densities above 9--10 T when cooled down to a physical temperature of 4.2 K. However, using more powerful, and demanding, cryogenic systems, NbTi can be cooled down to 2 K, which makes field strengths of up to about 11.7 T frequent \cite{6948313, Liebel2012, 2013SuScT..26i3001L}. Therefore, regular multicoil magnets and solenoids present field strengths of 9.4 or 11.7 T with a warm bore of about 50--90 cm, from a few to several meters in length, and a field homogeneity of several dozens parts per million over a few centimeters diameter of spherical volume. In occasions, niobium–tin,  $Nb_3$Sn, is used to provide field strengths beyond 11.7 T, with an upper limit of approximately 23.5 T. However, this superconductor material is, roughly, one order of magnitude more costly than NbTi and, therefore, its use is less widespread. Finally, high-temperature superconductors have been proposed to reach magnetic fields in excess of 23.5 T, although they are still at an early stage of development, while hybrid magnets that far surpass the state of the art are also maturing. Solenoids and multicoil superconducting magnets are employed in magnetic resonance imaging (MRI), industry and research \cite{6948313, Liebel2012, 2013SuScT..26i3001L}. DALI is designed to employ a regular MRI-type magnet with a high field stability and ultra high homogeneity, at the level of $<$1\% inhomogeneity over a 100 mm diameter sphere, thereby ensuring a constant field density in the volume destined to enclose the instrument; which can operate in multiple positions. An initial, cost-effective phase using a NbTi superconducting magnet operating at $\sim$4 K to provide a $\sim$9 T field is planned for the DALI Experiment, which we will refer to as `phase I.' A second phase, with a higher scan speed, is an up-grade to $\sim$11.7 T using a larger magnet working at $\sim$2 K. In both phases of the project only currently available technology is to be used as part of a strategy to contribute an experiment with feasibility and readiness. Note, the phase I configuration would also allow the search for DFSZ I axions up to $\sim$250 $\upmu$eV, although this would multiply by a factor of about 20 the time required to probe a band at a given confidence level.

\section{Discovery potential}\label{III}
In this section we look at the potential of DALI to unveil new physics.
\subsection{CP symmetry and axion dark matter}
The QCD Lagrangian density includes an angular term, which reads

\begin{equation}
\label{equation_1}
\pazocal{L}_{\theta}= \frac{1}{32\pi^2} \,\theta \, G^{a}_{\mu\nu} \tilde{G}^{a}_{\mu\nu} \, ,
\end{equation}
where the gluon field strength is $G^{a}_{\mu\nu}$ and $\tilde{G}^{a}_{\mu\nu}$ its dual. We express natural units throughout this letter, except where indicated explicitly. In the most natural explanation to date, CP symmetry in preserved in QCD since the $\theta$ term in Eq. \ref{equation_1}, which allows for symmetry violation, is promoted to a field with a Mexican hat potential that evolves towards a minimum as time advances in an early Universe \cite{PhysRevLett.38.1440}. Oscillations give rise to particles, which have been termed `axion' \cite{PhysRevLett.40.223, PhysRevLett.40.279}. The non-observation of a neutron electric dipole moment by modern experiments suggest that $\theta$ is negligibly small and, consequently, CP symmetry would be preserved, in practice, at present \cite{PhysRevLett.124.081803, PhysRevLett.97.131801}. Axion interacts weakly with Standard Model particles. The axion-photon interaction term is

\begin{eqnarray}
\pazocal{L}^{\mathrm{int}}_{a\gamma}=\label{Eq.1} -\frac{1}{4}g_{a\gamma\gamma} F_{\!\mu \nu} \tilde{F}^{\mu \nu}\!a  
\,,
\end{eqnarray}
with $F^{\mu \nu}$ the as photon field strength tensor and $a$ as the axion field. From a classic approach, Eq. \!\ref{Eq.1} simplifies to $\pazocal{L}^{\mathrm{int}}_{a\gamma}=g_{a\gamma\gamma}  \, \mathord{\mathrm{E}} \cdot \mathord{\mathrm{B}}\, a $; $\mathbf{\mathrm{E}}$ being the photon field and $\mathbf{\mathrm{B}}$ a static magnetic field that contributes a virtual photon that enables the Primakoff effect at an axion-photon-photon vertex, $a+\gamma_{\mathrm{virt}} \leftrightarrow \gamma$ \cite{Primakoff:1951iae}.

The coupling rate of the QCD axion contains a factor derived from the ratio of 
electromagnetic and color anomalies, $\pazocal{E}/\pazocal{C}$, which reads
 $c_{a\gamma\gamma}=1.92(4)- \pazocal{E}/\pazocal{C}$; with $c_{a\gamma\gamma}=-\frac{\upalpha}{2\pi}g_{a\gamma\gamma} f_{a} $, $\upalpha$ being the fine structure constant and $f_{a}$ the axion field scale—giving the digit in parentheses accounts for the uncertainty. In the Kim--Shifman--Vainshtein--Zakharov (KSVZ)
model,
$\pazocal{E}/\pazocal{C}=0$ and $\pazocal{C}=1$ are adopted \cite{PhysRevLett.43.103, Shifman1980CanCE}. In contrast, the Dine--Fischler--Srednicki--Zhitnitsky (DFSZ)
axion adopts $\pazocal{E}/\mathrm{\pazocal{C}}$ equal 8/3—the so-called DFSZ I—or 2/3—DFSZ II—and $\mathrm{\pazocal{C}}$ equal to 6 or 3 \cite{DINE1981199, osti_7063072}.
Differently from QCD axion models, for the so-called axion-like particles (ALPs), which arise in extensions of the Standard Model of particle physics, coupling to
photons and mass are uncoupled, resulting in a larger parameter space to be explored \cite{doi:10.1146/annurev-nucl-120720-031147, doi:10.1126/sciadv.abj3618}.

The core objective of DALI is to detect Galactic axion dark matter. The haloscope is sensitive to axion-like particles with a coupling strength to photons of \cite{De_Miguel_2021}
\begin{equation}
\begin{aligned}
\frac{g_{a \gamma\gamma}}{\mathrm{GeV^{-1}}}\!\gtrsim\!2.7\times10^{-13}  \times \left(\frac{\mathrm{SNR}}{Q}\right)^{1/2} \\ \!\! \times \left(\frac{\mathrm{m}^2}{A}\right)^{1/2}\!\! \times \left(\frac{m_{a}}{\mathrm{\upmu eV}}\right)^{5/4}
 \!\! \times \left(\frac{1\,\mathrm{s}}{t}\right)^{1/4}\\ \!\! \times \left(\frac{T_\mathrm{{sys}}}{\mathrm{K}}\right)^{1/2}  \!\! \times \frac{\mathrm{1\,T}}{B_0} \times \left(\frac{\mathrm {GeV cm^{-3}}}{\rho_{a}}\right)^{1/2} \, ,
\end{aligned}
\label{Eq.3}
\end{equation}
where SNR is signal to noise ratio, $Q$ is the quality factor, $A$ is cross-sectional area, $t$ is integration time, $T_\mathrm{{sys}}$ is the system temperature, $B_0$ is magnetic field strength and $\rho_{a}$ is the density of axion dark matter.

\begin{figure}[h]\centering
\includegraphics[width=.5\textwidth]{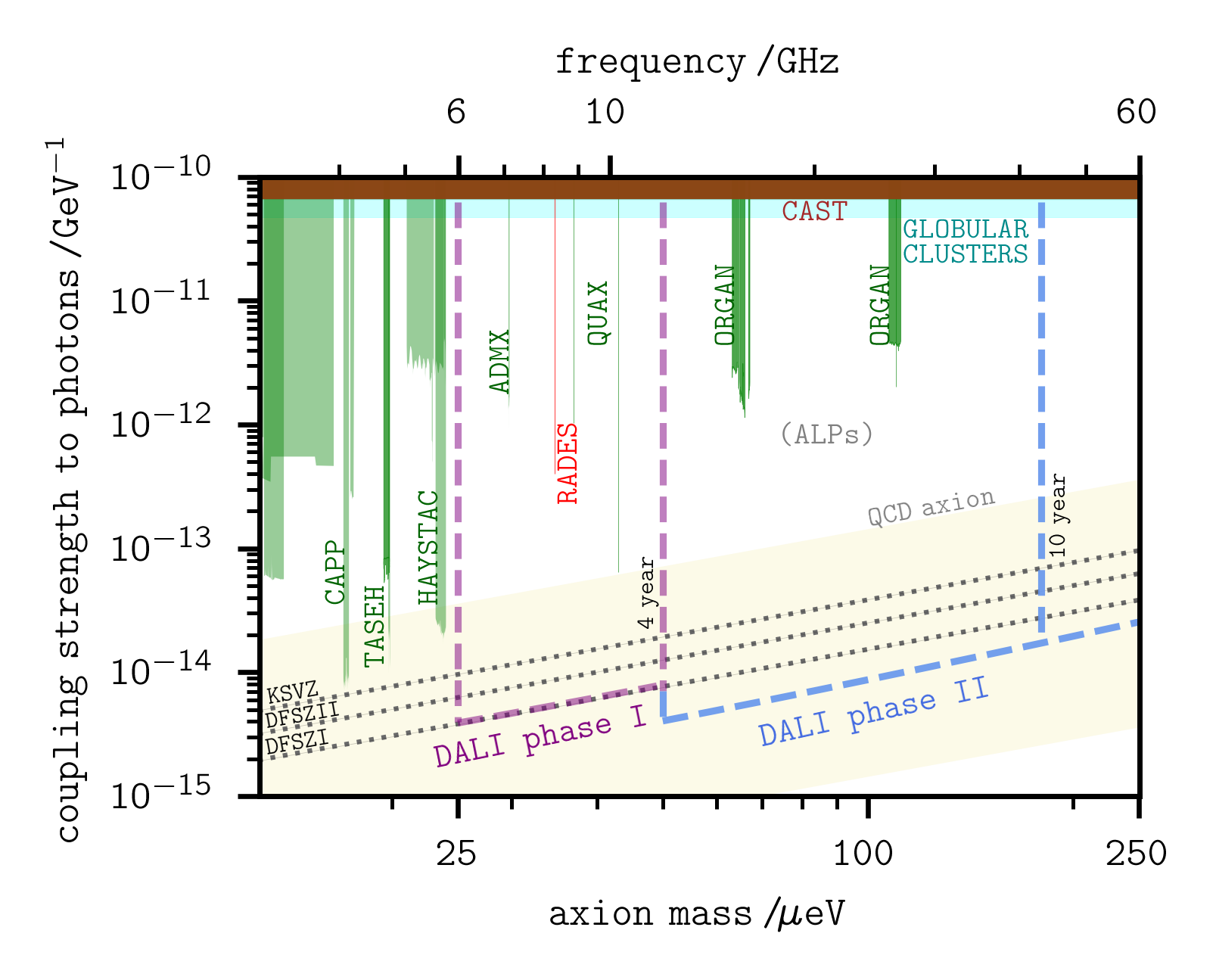}
\caption{A forecast of the sensitivity of DALI Experiment to Galactic axion dark matter projected onto current exclusion limits that are differentiated by color
\cite{PhysRevD.105.035022, PhysRevD.98.103015, Marsh_2017, PhysRevLett.118.011103, 2014PhRvL.113s1302A, Straniero:2015nvc, 2022JCAP...10..096D, 2022JCAP...02..035D, REGIS2021136075, 2021Natur.590..238B, EHRET2010149, PhysRevD.88.075014, 2016EPJC...76...24D, PhysRevD.92.092002,PhysRevLett.59.839, PhysRevD.42.1297, PhysRevLett.104.041301,PhysRevD.97.092001,PhysRevLett.128.241805,doi:10.1063/5.0098783,2021Natur.590..238B, CAST:2017uph, CAST:2020rlf, Foster:2020pgt, Darling:2020uyo, Darling:2020plz, PhysRevLett.121.261302, PhysRevD.99.101101, PhysRevD.103.102004, MCALLISTER201767, doi:10.1126/sciadv.abq3765, AxionLimits, Quiskamp:2023ehr}. The magnetic field is 9.4 T or 11.7 T for phase I, in purple, and phase II, in blue, respectively. The cross-sectional plate area is 1/2 m$^2$ or 3/2 m$^2$, and about four dozen layers are stacked in series to provide a high quality factor—cf. \cite{supp} for details. The system temperature is determined by a sub-kelvin background temperature provided by $^3$He coolers plus an offset contributed by the limit of heat dissipation in the electronics, roughly 2--3 times over the quantum noise limit, in order to be consistent with the frequency-dependent noise figure in high-electron-mobility transistor technology with a physical temperature at the level of one kelvin \cite{2017JATIS...3a4003M, PMID:25384166}. This causes a slope with respect to the QCD axion model lines that is partially compensated by a frequency-dependent quality factor. The instantaneous scanning bandwidth is between several dozens to a few hundred megahertz; while the axion-induced signal linewidth is $\Delta \nu/\nu\approx5\times10^{-7}$. The KSVZ and DFSZ axion models are projected over the entire experimental range, 25--250 $\upmu$eV. The QCD axion window is shaded in yellow \cite{DiLuzio:2016sbl}. The region in white is compatible with axion-like particles (ALPs).}
\label{fig_2}
\end{figure}

In the light of the sensitivity projections in Fig. \ref{fig_2}, DALI has a potential to probe DFSZ I-type axion models between 25--50 $\upmu$eV within approximately four years in the initial phase of the experiment. During phase II,  sensitivity will be increased allowing DALI to search for DFSZ I axion in the 50--180 $\upmu$eV over a period of about ten years.

We recompute the sensitivity of the DALI haloscope by means of Monte Carlo simulations in the Supplemental Material accompanying this manuscript \cite{supp}.

\subsection{Other purposes of the DALI project}
The detection or parametric constriction of the axion can shed light on a number of problems in physics that are reviewed in the following paragraphs.
\subsubsection{Examining cosmology in a post-inflationary Universe}
The color anomaly is referred to as `domain wall number,' $\pazocal{N}\leftarrow\pazocal{C}$, in cosmology. The moment in the history of the early Universe in which the axion angular field, $\theta$, acquires propagating degrees of freedom is called the `phase transition.' In an scenario in which the phase transition takes place before inflation, axion strings, domain walls, emerging if $\pazocal{N}>1$, and their remnants, will be cleaned out by the expansion. If symmetry breaking originates after inflation, then domain walls would invoke catastrophic topological objects \cite{Linde:1991km, Hiramatsu:2012gg, MARSH20161}. For axion models that adopt $\pazocal{N}=1$, such as the KSVZ model, these topological defects are naturally avoided. In a post-inflationary scenario, models with $\pazocal{N}>1$, including the DFSZ
axion, must circumvent those dramatic topologicals by alternative mechanisms \cite{Barr:2014vva, PhysRevD.98.043535, PhysRevD.93.085031, hiramatsu2011evolution, kawasaki2013domain, harigaya2018qcd}.

Information can be extracted from this picture. The axion mass in a post-inflationary scenario is often restricted to the range $25 \, \upmu \mathrm{eV} \lesssim m_{a} \lesssim  1 \,  \mathord{\mathrm{ meV }}$—cf. \cite{MARSH20161, PhysRevD.98.030001, PhysRevD.105.055025}. Recent studies even constrain this range further, to $40\lesssim m_{a}/\mathrm{\upmu eV}\!\lesssim\!180$, to which DALI will pay special attention \cite{2022NatCo..13.1049B}.

It is equally appealing that DALI will simultaneously explore the axion mass range constrained by models that unify the strong 
CP problem, dark matter, usually adopting a KSVZ-type axion, and cosmic inflation, simultaneously solving other problems of modern physics. This band is  $50\lesssim m_{a}/\mathrm{\upmu eV}\!\lesssim\!200$ in order to be compatible with observational data and the $\Lambda$CDM cosmology  \cite{Salvio:2015cja, ballesteros2017standard, ballesteros2017unifying, ernst2018axion, 2019FrASS...6...55R, ballesteros2017unifying}.
\subsubsection{Astrophysical bounds of axion dark matter}
The confrontation of stellar evolution models, modified to account for the additional energy loss rate that axion scattering would cause, with observational data, allows to set limits to the coupling constants of the pseudoscalar with ordinary particles. Axion-like particles coupling to both photons and fermions could simultaneously explain both the observed extra cooling in horizontal branch stars, which is compatible with the Primakoff effect, and the abnormally accelerated evolution of red giant branch stars that may originate from axion coupling to electrons in several radiative processes such as atomic axio-recombination 
or deexcitaction, axion bremsstrahlung and Compton scattering. This could also explain additional cooling in white dwarfs through axion-induced losses due to bremsstrahlung; all in concordance with the discrepancy of the neutrino flux duration from supernova (SN) SN1987A compared to simulations, and some intriguing observations of the neutron star in the SN remnant Cassiopeia A that disclose an 
abnormally fast cooling rate compatible with axion-neutron bremsstrahlung \cite{PhysRevD.36.2211, Raffelt:1999tx, Viaux:2013lha, Ayala:2014pea, MillerBertolami:2014rka, Fischer:2016cyd, Carenza:2019pxu, Leinson:2014ioa, Keller:2012yr, Giannotti:2017hny, Straniero:2020iyi, diluzio2021stellar, 2022JCAP...10..096D}. This fosters the exploration of the corresponding mass range with DFSZ axion sensitivity. DALI has a potential to probe these astro-bounds, whose most restrictive limits in its band to date, resulting from observations of globular clusters, are projected onto Fig. \ref{fig_2} \cite{Ayala:2014pea, 2022JCAP...10..096D}, with DFSZ I sensitivity.
 
\subsubsection{Dark sector photons}
Dark photon, also referred to as hidden photon or paraphoton, is a hypothetical gauge boson that mixes kinetically with ordinary photons \cite{Okun:1982xi, Vilenkin:1986}. The interaction term relevant for this work reads
\begin{equation}
\pazocal{L}^{\mathrm{int}}_{\gamma'\gamma}= 
-{\frac {1}{2}}  F_{\mu {\nu} } \tilde{X}^{\mu \nu } \chi \,,
\label{Eq.9}
\end{equation}
where we denote by $\tilde{X}^{\mu \nu}$ the field strength tensor of the dark photon field;  $\chi$ being the dimensionless kinetic mixing strength.
\begin{figure}[h]\centering
\includegraphics[width=.5\textwidth]{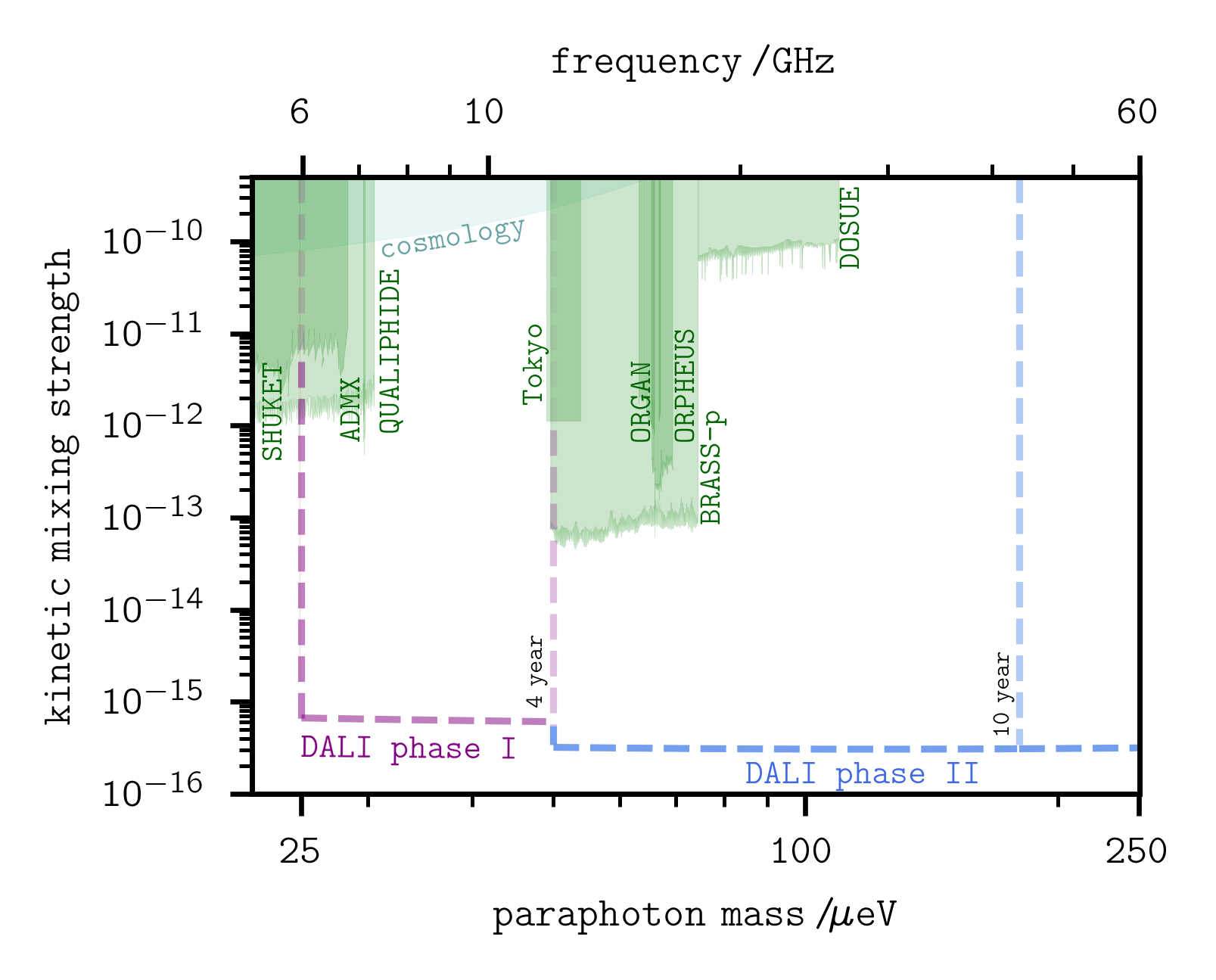}
\caption{Projection of the sensitivity of DALI to dark sector photon dark matter with overlapping boundaries published at the time of writing \cite{Arias:2012az, Suzuki:2015vka, ADMX:2018ogs, Brun:2019kak,  Dixit:2020ymh, Cervantes:2022epl, Cervantes:2022yzp, Bajjali:2023uis, McAllister:2022ibe, DOSUE-RR:2022ise, PhysRevLett.130.231001}. Sensitivity  enhancement in a fixed polarisation scenario owing to an unrestricted directionality, as suggested in \cite{Caputo:2021eaa}, is not rescaled in this plot.}
\label{fig_5}
\end{figure}

The sensitivity of DALI to Galactic paraphotons is

\begin{equation}
\begin{aligned}
\chi \gtrsim 2.9\times10^{-14}  \times \left(\frac{\mathrm{SNR}}{Q}\right)^{1/2}  \!\! \times \left(\frac{\mathrm{m}^2}{A}\right)^{1/2}\!\! \times \left(\frac{\mathrm{\Delta\nu}}{\mathrm{Hz}}\right)^{1/4}
 \\ \times  \left(\frac{\mathrm{1\,s}}{t}\right)^{1/4} \!\! \times \left(\frac{T_\mathrm{{sys}}}{\mathrm{K}}\right)^{1/2}  \!\!  \times \left(\frac{\mathrm {GeV cm^{-3}}}{\mathrm{\rho_{\gamma\prime}}}\right)^{1/2} \!\!\times \frac{\sqrt{2/3}}{\alpha}\, ,
\end{aligned}
\end{equation}
$\alpha=\sqrt{2/3}$ representing random incidence angle of the dark photon \cite{Horns:2012jf}. Figure \ref{fig_5} shows a forecast of DALI sensitivity considering that hidden photons make up a large part of the Galactic dark matter.

\subsubsection{Exploration of a dark universe}
\begin{figure*}[]\centering
\includegraphics[width=0.9\textwidth]{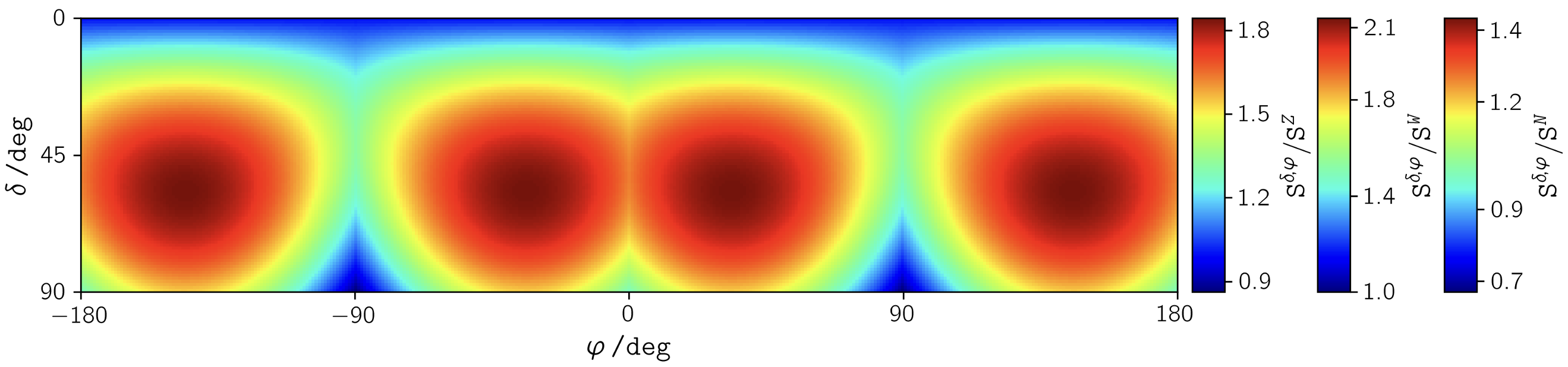}
\caption{Significance of the signal modulation, $\mathrm{S}^{\delta,\varphi}$, for a collision with a substructure compared to a zenith, north or west pointing experiment, $\mathrm{S}^{Z, N, W}$. The inclination angle, $\delta$, is measured from the zenith direction in lab coordinates; and $\varphi$ is the azimuth angle of its orthogonal projection on a north-west plane measured from the north fixed reference. The optimal pointing varies with lab coordinates, and smoothly over time. This simulation is independent from the axion model and mass and the characteristic properties of the axion flow—density, velocity and dispersion, etc.}
\label{fig_3}
\end{figure*}

Different patches, causally disconnected as a result of a scenario at which the phase transition takes place after cosmic inflation, could favor the 
formation of substructures of axion dark matter \cite{Hogan:1988mp}. If a substructure were to traverse our planet, it would induce a measurable imprint on the electromagnetic spectrum. Substructures have been shown to approach the Solar System by analyzing survey data, with an event rate that may not be insignificant \cite{o2017axion,2018JCAP...11..051K,PhysRevD.98.103006}. The daily signal modulation can be expressed in terms of $c_0, c_1$ and $\phi$, in the form $c_0+c_1\mathord{\mathrm{cos}}(2\pi t/0.997+\phi)$; $t$ being the time measured in days from January 1st, and

\begin{subequations}
\begin{eqnarray}
 \overbracket{b_0\,\mathord{\mathrm{cos}}\lambda_{\mathord{\mathrm{lab}}}}^{c_0^{{N}}}- \overbracket{b_1\, \mathord{\mathrm{sin}}\lambda_{\mathord{\mathrm{lab}}}}^{c_1^{N}}\mathord{\mathrm{cos}}(\omega_d\,t+ \overbracket{\phi_{\mathord{\mathrm{lab}}}+\psi}^{\;\:\phi^{ {N}}}\,)
\,,
\\
\overbracket{b_1}^{{c_1^{ {W}}}}  \mathord{\mathrm{cos}}(\omega_d\,t+ \overbracket{\phi_{\mathord{\mathrm{lab}}}+\psi-\pi}^{\;\:\phi^{ {W}}}\,)
\,,\\
 \overbracket{b_0\,\mathord{\mathrm{sin}}\lambda_{\mathord{\mathrm{lab}}}}^{c_0^{ {Z}}}+ \overbracket{b_1 \,\mathord{\mathrm{cos}}\lambda_{\mathord{\mathrm{lab}}}}^{c_1^{ {Z}}}\mathord{\mathrm{cos}}(\omega_d\,t+ \overbracket{\phi_{\mathord{\mathrm{lab}}}+\psi}^{\;\:\phi^{ {Z}}}\,)
\,,
\label{Eq.17}
\end{eqnarray}
\end{subequations}
where  $b_0=\sigma_3\mathord{\mid\!\!\mathord{\mathrm{V\!_{lab}}}-\mathord{\mathrm{V}}\!_{a}\!\!\mid}^{-1}$, $b_1=(\sigma_1^2+\sigma_2^2)^{1/2}\mathord{\mid\!\!\mathord{\mathrm{V\!_{lab}}}-\mathord{\mathrm{V}}\!_{a}\!\!\mid}^{-1}$, $\psi=\mathord{\mathrm{tan}}^{-1}(\sigma_1/\sigma_2)-0.721\omega_d-\pi/2$; $\sigma_1=(-0.055,0.494,-0.868)\cdot \Upsilon$, $\sigma_2=(-0.873,-0.445,-0.198)\cdot \Upsilon$, $\sigma_3=(-0.484,0.747,0.456)\cdot \Upsilon$, with $\Upsilon=\mathord{\mathrm{V_{\odot}}}+v_{\oplus}\,\mathord{\mathrm{cos}}(\tau_y)(0.994, 0.109, 0.003)+v_{\oplus}\,\mathord{\mathrm{sin}}(\tau_y)(-0.052, 0.494, -0.868)$,  $\tau _y=2 \pi \, (t-79)/ 365$ \cite{2018JCAP...11..051K, De_Miguel_2021}. 

In the particular case of the axion, which has no polarization, the directionality originates along the dimensions in which the detector has a fraction of at least 1/5 of the de Broglie wavelength \cite{2018JCAP...11..051K}. DALI incorporates an altazimuth mount which enhances its sensitivity to dark matter flows. This is particularly interesting as DALI
is envisioned to probe the post-inflationary axion, in a scenario at which topological defects may give rise to substructures. The simulation of an event is shown in Fig. \ref{fig_3}. Interestingly, the search for Halo axions and substrcutures can be performed simultaneously, analyzing the data in parallel. Detection of this trace would support the hypothesis of substructures navigating in a `dark universe' \cite{HOGAN1988228, moore1999dark, diemand2008clumps,2011MNRAS.413.1419V,2016JCAP...01..035T,2018MNRAS.475.1537M,vaquero2019early}.

\subsection{Peripheral objectives of the project}
\subsubsection{Probing axion quark nuggets}
It has been suggested that the phase transition could concentrate most of the quark excess in the form of invisible quark nuggets, thereby explaining dark matter in the QCD framework \cite{PhysRevD.30.272}. This could give raise to the observed density ratio between non-luminous and visible matter, $\Omega_{\mathrm{dark}}/ \Omega_{\mathrm{visible}}\sim1$. This hypothesis has been transferred to axion \cite{Liang:2016tqc, Fischer:2018niu, Liang:2018ecs, Liang:2019lya, Budker:2019zka, Zhitnitsky:2021iwg}. Axion quark nuggets (AQNs) are dense relic specks that could arise regardless of the initial misalignment angle or axion mass in a $\pazocal{N}=1$ scenario. The sensitivity of a DALI-like device to axions released by AQNs is read

\begin{equation}
\begin{aligned}
\frac{g_{a \gamma\gamma}}{\mathrm{GeV^{-1}}}\!\gtrsim\!2.1\times10^{-7}  \times \left(\frac{\mathrm{SNR}}{Q(v_{a})}\right)^{1/2} \!\! \times \left(\frac{\mathrm{m}^2}{A_{s}}\right)^{1/2} \\ \!\! \times \left(\frac{m_{a}}{\mathrm{\upmu eV}}\right)^{5/4}
 \!\! \times \left(\frac{\mathrm{1\,s}}{t}\right)^{1/4} \!\! \times \left(\frac{T_\mathrm{{sys}}}{\mathrm{K}}\right)^{1/2}  \!\! \times \frac{\mathrm{1\,T}}{B_0} \\ \times \left(\frac{\mathrm {eV cm^{-3}}}{\rho^{\mathrm{AQN}}_{v_{a}}}\right)^{1/2} \!\! \times\left(\frac{v_{a}}{c}\right)^{1/2} \!\! \times \frac{\mathrm{1}}{\sqrt{n}} \, .
\end{aligned}
\label{Eq.10}
\end{equation}

A scaled-down parasitic detector with the same concept as DALI can be devoted to the exploration of the axion flux induced by AQNs on Earth. Such an array of $n$ independent pixels would have a smaller plate scale, $s$, of the order of a dozen centimeters, in order to maintain de Broglie coherence. This would give the prototype access to relativistic axions of up to, roughly, 100 $\upmu$eV of mass \cite{De_Miguel_2021}; with a cut-off close to $c/5$—note that the spectral density function of AQN-induced axions peaks at about $c/2$. From \cite{Liang:2019lya, Budker:2019zka}, it follows that the fluence of semi-relativistic axions on Earth originating from collisions with those macroscopic specs is $\Phi^{\mathrm{AQN}}_{v_{a}<c/5}\sim10^{12}\mathrm{cm^{-2}s^{-1}}(\mathrm{eV}/m_{a})$, which results in an occupancy of about $\rho^{\mathrm{AQN}}_{v_{a}<c/5}\sim10^{2}$  $\mathrm{eVcm}^{-3}$. That is several orders of magnitude less concentrated than the saturation density of dark matter at the position of the Solar System, about few$\times10^8$ $\mathrm{eVcm}^{-3}$. However, recent work has pointed out that a gravitationally focused stream of AQNs could transiently increase the occurrence of annihilation events by a factor of up to $10^6$ \cite{Patla_2014}. Notwithstanding that the semi-relativistic velocities of those axions, $v_{a}\lesssim c/5$, would decrease the signal boost factor \cite{Millar:2017eoc}, a partially resonant, not purely transparent, harmonically hybrid mode could take advantage of the larger momentum transmitted by the rapid axions, allowing to maintain, or perhaps increase, the enhancement factor, $Q(v_{a})$, with respect to the low-velocity limit. In addition, signal modulation caused by celestial mechanics, and the so-called `local-flashes,' bursts resulting from the interaction of AQNs with the Earth in the vicinity of a detector, could result in an amplification parameter with magnitude $10^{2-4}$ during a short period of time with a non-negligible event rate \cite{Liang:2019lya, Budker:2019zka}. The sensitivity is multiplied by $[(v_{a}/c)^2/10^{-6}]^{1/4}$ on the right-hand side of Eq. \ref{Eq.10} to give account of the broadening and subsequent dilution of the signal in frequency domain caused by the semi-relativistic velocity of the AQN-induced axions compared to that of virialized particles. However, the above aspects, separately or together, may give the instrument access to the KSVZ axion window. 

Equation \ref{Eq.10} can be transferred to any exotic source of semi-relativistic axion astroparticles. One example could be the axion-compatible explanation for the intriguing Antarctic Impulse Transient Antenna events \cite{PhysRevLett.117.071101, ANITA:2018sgj, Esteban:2019hcm}.
\subsubsection{Constraints on the diffuse gravitational wave background}

Graviton, the gravitational wave (GW) counterpart in the form of a long-postulated elementary particle, mix with photons in the presence of static magnetic fields \cite{Gertsenshtein, PhysRevD.37.1237, Dolgov:2012be}. The Lagrangian density of the interaction is 

\begin{equation}
\pazocal{L}^{\mathrm{int}}_{g\gamma}=-\frac{1}{2}\kappa\,h_{\mu\nu}\, \mathord{\mathrm{B}^{\mu}}\, \mathord{\mathrm{B}^{\nu}} \;,
\label{Eq.99}
\end{equation}
where the gravitational coupling is $\kappa^2=16\pi m_{\mathrm{Pl}}^{-2}$, with $m_{\mathrm{Pl}}\sim10^{19}$ GeV the Planck mass; $h_{\mu\nu}$ denotes the graviton, $\mathord{\mathrm{B}^\mu}$ is the external magnetic field and $\mathord{\mathrm{B}^\nu}$ the electromagnetic wave \cite{Cembranos:2018jxo}.

Irreducible emission from the time evolution of the momentum-energy tensor and cosmic string decay, the coalescence of primordial black hole binaries and the evaporation of low-mass primordial black holes, branes oscillation, or those of astrophysical origin, such as solar thermal gravitational waves, supernova collapse, rotating neutron stars, binary systems, etc., contribute to shape a diffuse gravitational wave background (GWB) \cite{Marck:1997re, Maggiore:1999vm, Servin:2003cf, Bisnovatyi-Kogan:2004cdg, Clarkson:2006pq, Ejlli:2019bqj}.

\begin{figure}[h]\centering
\includegraphics[width=.5\textwidth]{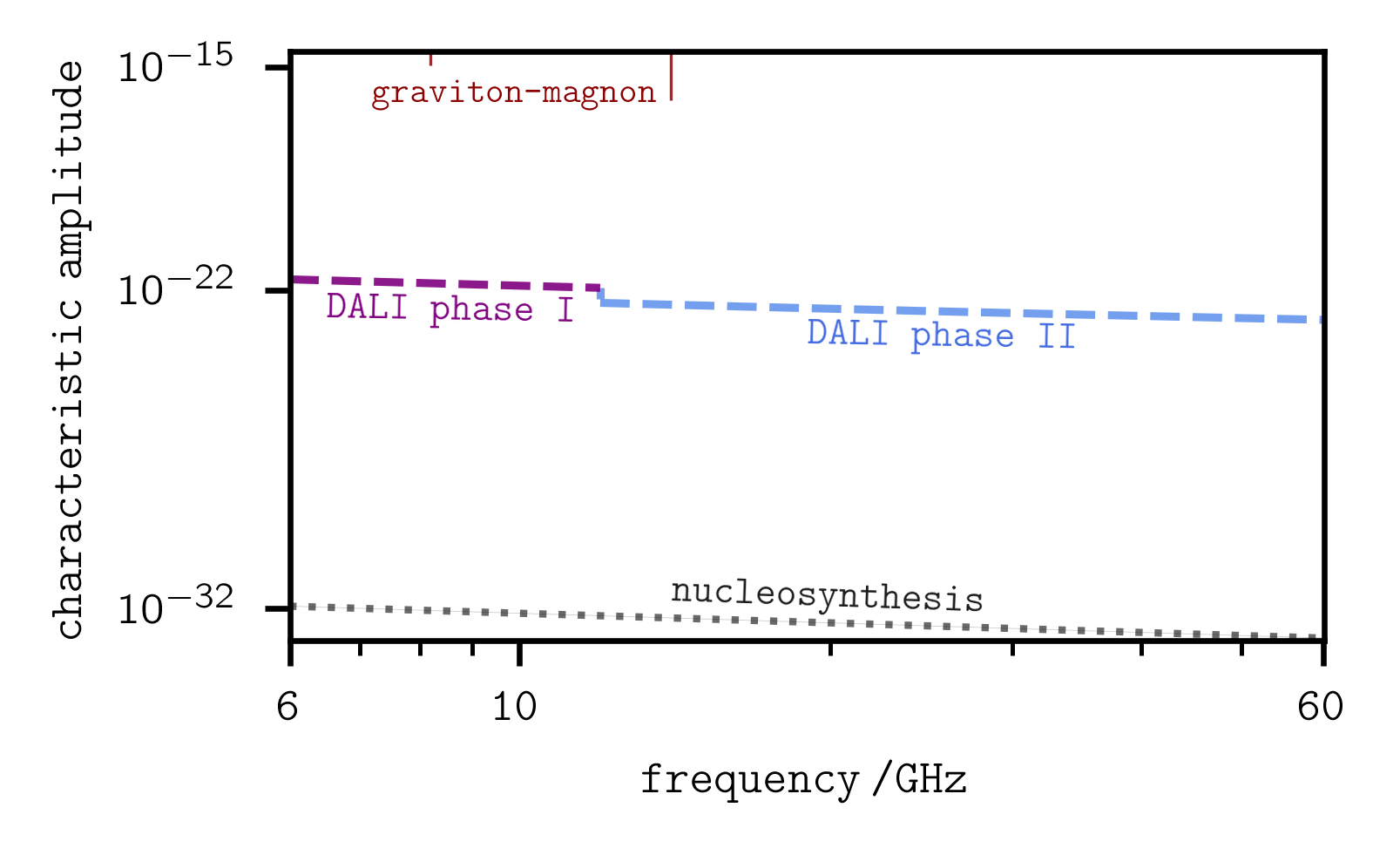}
\caption{Accessible characteristic amplitude of a high-frequency stochastic and isotropic
gravitational wave background. Results overlapping the experimental range, from \cite{Ito:2019wcb}, are plotted in red.}
\label{fig_4}
\end{figure}

In the spirit of \cite{Maggiore:1999vm}, we now introduce the dimensionless characteristic amplitude $h_c=(6\,\Omega_{\mathrm{GW}})^{1/2}H_0\,\omega^{-1}$; with $\Omega_{\mathrm{GW}}(\omega)$ as the spectral density function of a stochastic GWB, $H_0$ as the Hubble parameter, and $\omega$ as the pulse of the wave. Primordial gravitational waves contribute to the density of species during nucleosynthesis as massless neutrinos, which results in a higher freezing temperature at which expansion breaks the $p e \leftrightarrow n\nu_e$ equilibrium. This affects the baryon to photon ratio and constraints $h_c \lesssim 4.5\times10^{-22}\,\mathrm{Hz}/\omega$. 

Unfortunately, the cross-section of the photon-graviton oscillation in a magnetic field is weighted by $m_{\mathrm{Pl}}^{-2}$. Therefore, the direct detection of the high-frequency GWB through this approach cannot be tackled with current technology. In any case, some experiments have used the weak coupling of the graviton to ordinary particles to set bounds to the amplitude of the GWB at shorter wavelengths. Interferometers have established $h_c \gtrsim 10^{-18}$ at 1 MHz, $h_c \gtrsim 10^{-19}$ at 13 MHz and $h_c\gtrsim 10^{-10}-10^{-12}$ at 100 MHz \cite{2017PhRvD..95f3002C, Cruise:2006zt, Nishizawa:2007tn, Akutsu:2008qv}; while the graviton-magnon resonance detector in \cite{Ito:2019wcb} establishes $h_c \gtrsim 10^{-15}$ at around 8 GHz and $h_c \gtrsim 10^{-16}$ at 14 GHz. The analysis carried out in \cite{Ejlli:2019bqj} using data from light-shining-through-walls experiments in \cite{OSQAR:2015qdv, Ehret:2010mh}, and helioscopes \cite{CAST:2017uph}, establishes the upper bounds $h_c \gtrsim 10^{-25}$ and  $h_c \gtrsim 10^{-25}$ at $10^{14-15}$ Hz, and $h_c \gtrsim 10^{-27}$ at approximately $10^{18}$ Hz. 

DALI may set the most restrictive experimental constraints to the amplitude of the high-frequency GWB in its band via graviton-to-photon conversion in its magnetised vessel. Following the approach in \cite{Ejlli:2019bqj, doi:10.1063/1.1770483}, it is possible to infer the minimum amplitude of a stochastic GWB that can be measured by the experiment \begin{equation}
\begin{aligned}
h_c\gtrsim9.2\times10^{-12}  \times \left(\frac{\mathrm{SNR}}{Q}\right)^{1/2}  \!\! \times \left(\frac{\mathrm{m}^2}{A}\right)^{1/2}\!\! \times \left(\frac{\mathrm{Hz}}{\nu}\right)^{1/2}
 \\ \times \left(\frac{\mathrm{1 \,s}}{t}\right)^{1/4} \times\left(\frac{\mathrm{Hz}}{\mathrm{\Delta\nu}}\right)^{1/4} \times \left(\frac{T_\mathrm{{sys}}}{\mathrm{K}}\right)^{1/2}  \!\! \times \frac{\mathrm{1\,T}}{B_0} \times \frac{1\,\mathrm{m}}{L} \, ,
\end{aligned}
\end{equation}
where $L$ is the length of flight.  A sensitivity projection is shown in 
Fig. \ref{fig_4}.

\section{Conclusions}\label{IV}
\begin{table*}[t]
\caption{Discovery potential of DALI. Recap of the core and peripheral project goals discussed throughout this paper. Axion detection refers to the first third, dark photon physics to the second inset, upper limits to a gravitational wave background follow the last horizontal line. The sensitivity is expressed in terms of ${g_{a\gamma\gamma}}$ by means of a reference axion model, KSVZ or DFSZ; the kinetic mixing constant, $\chi$, for the paraphoton or the characteristic amplitude of a stochastic gravitational wave background, $h_c$, for the graviton. The potential impacts of this research are noted.}
\begin{ruledtabular}
\begin{tabular}{cccc}
\label{table1}

\textrm{Branch and purpose}&
\textrm{Band}&
\textrm{Sensitivity }&\textrm{Impact}\\
\colrule \rule{0pt}{2.5ex}Charge\&parity (CP) symmetry & 25--250 $\upmu$eV & DFSZ I&Strong CP problem in quantum chromodynamics washed\footnote{Quantum chromodynamics solution to the so-called strong CP problem; a new pseudo-scalar added after the Higgs boson \cite{ATLAS:2012yve, CMS:2012qbp, CMS:2013btf}.} \\
Galactic axion-like dark matter & 25--250 $\upmu$eV  & DFSZ I& Direct detection of dark matter; physics realignment\footnote{If ALPs are found, then a significant part of dark matter, cosmology, and halo modeling, among other issues, may undergo fine adjustment.}\\
$\Lambda$CDM cosmology (inflation) &  40--180 $\upmu$eV &  DFSZ I & Heavier axion sustains cosmic inflation\footnote{Observation of axion of mass heavier than several dozens of microelectronvolts would support a post-inflationary scenario.}\\SMASH-type cosmology &  50--200 $\upmu$eV &  >KSVZ & Standard model axion seesaw Higgs portal inflation (SMASH) test\footnote{The SMASH window may add three right-handed neutrinos, one Dirac fermion, a new
complex singlet scalar \cite{ballesteros2017unifying}.}\\
Dark universe (substructures) &  25--250 $\upmu$eV & DFSZ I &  Dark matter agglomerates anisotropically in macroscopic structures\footnote{Detection of a daily signal modulation can involve large substructures, affecting cosmology, large-scale structure, clusters, galaxies, etc.}\\
Stellar evolution (cooling rate) & 25--250 $\upmu$eV & DFSZ I& Fine-tuning of star model (HB, RGB, WDs, ...)\footnote{Confirmation of axion would allow fine-tuning of helium consumption rates; and conversely \cite{Ayala:2014pea, 2022JCAP...10..096D}.} \\
Axion quark nugget hypothesis &  25--80 $\upmu$eV &  KSVZ & Dark matter detection; domain wall number set $\pazocal{N}=1$\footnote{AQNs set domain wall number, suggests on decay of topologicals, solar ultraviolet radiation excess, `primordial lithium puzzle' \cite{Budker:2019zka}.}\\
\colrule \rule{0pt}{2.5ex}
Dark photon dark matter & 25--250 $\upmu$eV & $\chi\!\!\gtrsim\!\!5\!\!\times\!\!10^{-16}$ & Dark matter detection; Standard Model of particle physics extends\footnote{If this alternative candidate for dark matter is detected, a dark sector would be added to the Standard Model of particle physics.}\\
\colrule \rule{0pt}{2.5ex}

Gravitational wave background & 6--60 GHz  & $h_c\!\gtrsim\!\!10^{-22}$ &  Bounds to big bang nucleosynthesis by a direct detection method\footnote{More incisive experimental limits to the stochastic gravitational wave background provide useful information about the early Universe.}\\
\end{tabular}
\end{ruledtabular}
\end{table*}

The physics potential of DALI is summarised in Table \ref{table1}. By ramping the magnet on/off it would be possible to differentiate between axion-like and dark photon dark matter. On the other hand, the contribution of nucleosynthesis to an isotropic and stochastic gravitational wave background would manifest itself as a signal that cannot be detuned in small frequency steps. The collision with dark matter substructures would give raise to a peculiar modulation and a shortened signal duration. The spectral feature originated by AQN-induced axions may also have characteristics that would allow it to be discriminated.


\section*{Acknowledgements}
The work of J.D.M. was supported by RIKEN's program for Special Postdoctoral Researchers (SPDR)—Project Code: 202101061013—; J.F.H.C. is supported by the Resident Astrophysicist Programme of the Instituto de Astrofísica de Canarias. We gratefully acknowledge financial support from the Severo Ochoa Program for Technological Projects and Major Surveys 2020-2023 under Grant No. CEX2019-000920-S; Recovery, Transformation and Resiliency Plan of Spanish Government under Grant No. C17.I02.CIENCIA.P5; FEDER operational programme under Grant No. EQC2019-006548-P; IAC Plan de Actuación 2022. Actuación 2022. J.A.R.M. acknowledges financial support from the Spanish Ministry of Science and Innovation (MICINN) under the project PID2020-120514GB-I00. We thank R. Rebolo, H. Lorenzo-Hernández, R. Hoyland, J. Martín-Camalich, K. Zioutas and A. Zhitnitsky for discussions. This research made use of computing time available on the high-performance computing systems at the Instituto de Astrofísica de Canarias. The authors thankfully acknowledges the technical expertise and assistance provided by the Spanish Supercomputing Network (Red Española de Supercomputación), as well as the computer resources used: the deimos-diva supercomputer, located at the Instituto de Astrofísica de Canarias.
\bibliography{apssamp}

\clearpage
\pagebreak
\widetext
\begin{center}
\textbf{\large Discovery prospects with the Dark-photons \\ \& Axion-Like particles Interferometer}\\\vspace{5pt}
\textit{\large Supplementary Material}\\\vspace{8pt}
Javier De Miguel, Juan F. Hernández-Cabrera, Elvio Hernández-Suárez, 
Enrique Joven-Álvarez, Chiko Otani, and J. Alberto Rubiño-Martín
\end{center}
\setcounter{equation}{0}
\setcounter{figure}{0}
\setcounter{table}{0}
\setcounter{page}{1}
\setcounter{section}{0}
\makeatletter
\renewcommand{\theequation}{S\arabic{equation}}
\renewcommand{\thefigure}{S\arabic{figure}}
\renewcommand{\bibnumfmt}[1]{[S#1]}






\section{On the sensitivity projection to Galactic axions}
In this supplementary material we recompute the sensitivity of DALI to Galactic axion dark matter by means of a Monte Carlo simulation, for which we will mainly follow \cite{PhysRevD.96.123008, doi:10.1126/sciadv.abq3765}. The input data consists of $M$ individual spectra containing $2^N$ bins each generated synthetically. First, an average baseline is estimated by applying a Savitzky--Golay (SG) filter with a window size $W = 3001$ and a polynomial degree $d = 2$ to the average of all spectra. The average spectrum is then divided by the estimated baseline. Bins potentially compromised by intermediate frequency (IF) interferences are identified as any bin deviating more than $4.5 \sigma$ from the average as well as 3 bins to each side; the values of these bins are iteratively replaced by randomly generated values drawn from a Gaussian distribution with the same average and standard deviation as the values of the average spectrum until no further compromised bins are identified. The corresponding bins of the individual spectra are also replaced with random values in the same manner. Bins potentially compromised by spurious radio frequency (RF) interference are identified in individual spectra as any bin deviating more than several sigma from the average, the threshold being an arbitrary value that is established based on the exclusion sector to be considered. Their values are iteratively replaced by randomly generated values drawn from a Gaussian distribution until no further RF interferences are found. The spectra are individually normalised to their estimated baseline, which is obtained by means of a SG filter with the same parameters as before, to get the processed spectra. Processed spectra are now rescaled by the expected power boost at each frequency to calculate the rescaled spectra, namely 

\begin{equation}
            \delta_{ij}^s = \frac{\delta_{ij}^p}{Q_{ij}} \, ,
          \end{equation}

where $\delta$ is the value of a bin, $i$ is the index of a spectrum among all spectra, $j$ is the index of a bin within a spectrum and $Q_{ij}$ is the power boost provided by the resonator at each bin. The superscripts $s$ and $p$ represent the rescaled and the processed spectra, respectively. The standard deviation of each bin is given by

\begin{equation}
            \sigma_{ij}^s = \frac{\sigma_{i}^p}{Q{ij}} \, .
          \end{equation}

The rescaled spectra are then combined into a stacked spectrum. The RF bins corresponding to a single IF bin—i.e., non-overlapping bands—are copied directly onto the stacked spectrum; RF bins in overlapping bands are combined by means of a maximum likelihood (ML) weighted sum where the weights $w_{ijk}$ are calculated as

\begin{equation}
            w_{ijk} = \frac{\left (\sigma_{ij}^s \right )^{-2}}{\sum_{i'} \sum_{j'} \left (\sigma_{i' j'}^s \right )^{-2}} \, ,
          \end{equation}

where $k$ is the index of the stacked spectrum. The prime notation in the denominator is to be interpreted as a sum across all pairs $i,j$ corresponding to a bin $k$. The stacked spectrum can then be processed as

\begin{equation}
            \delta_k^c =  \sum_{i'} \sum_{j'} w_{ijk} \delta_{ij}^s \,
          \end{equation}

in the overlapping bands. The standard deviation of each stacked spectrum bin is calculated as

\begin{equation}
            \sigma_k^c =  \sqrt{\sum_{i'} \sum_{j'} w_{ijk}^2 \left ( \delta_{ij}^s \right )^2} \, .
          \end{equation}

Now the rebinned spectrum is obtained by merging neighbouring bins of the stacked spectrum in non-overlapping segments of a length equal to $K_r$ bins. Let $D_k^c = \delta_k^c (\sigma_k^c)^{-2}$ and $R_k^c = (\sigma_k^c)^{-1}$. The rebinned spectrum is obtained from 

\begin{equation}
             D_\ell^r = \sum_{k = (\ell - 1) K_r + 1}^{k = \ell K_r} D_k^c \,
          \end{equation}

and

\begin{equation}
            R_\ell^r = \sqrt{\sum_{k = (\ell - 1) K_r + 1}^{k = \ell K_r} (R_k^c)^2} \, ,
          \end{equation}

where $\ell$ is the index of the rebinned spectrum. The grand spectrum is obtained from a ML weighted sum taking into account the expected power spectral density or line shape $S_a(\nu)$ of the axion-induced signal. In this implementation, we have considered

\begin{equation}
          S_a \left ( \nu \right ) 
        = K \sqrt{\nu - \nu_a} \exp{\left (- \frac{3 (\nu - \nu_a)}{\nu_a \frac{\langle v_a^2 \rangle}{c^2}} \right )} \,,
          \end{equation}

at which $K$ is a constant added to scale $S_a(\nu)$ to its expected power. The weights are calculated by integrating over segments of $S_a(\nu)$ with a length of $K_r$ bins given a misalignment $\delta \nu_r$, according to 

\begin{equation}
\label{eq:PSD_axion}
           L_q \left ( \delta \nu_r \right ) =
        K_g \int_{\nu_a + \delta \nu_r + (q-1) K_r \Delta \nu_b}^{\nu_a + \delta \nu_r + q K_r \Delta \nu_b} S_a(\nu) \mathrm{d} \nu \,,
\end{equation}

where $q$ is an index running from 1 up to $K_g$—the length of the grand spectrum ML sum—and $\Delta \nu_b$ is the bin width of the original spectra. It should be noted that the lengths $K_r$ and $K_g$ must be chosen such that the axion width, $\Delta \nu$, is, approximately, $K_r K_g$ bins wide. The spectral density function $S_a (\nu)$ has a full width half maximum of about $5 \times 10^{-7} \nu_a$ if a value of (270 km/s$)^2$ is considered for $\langle v_a^2 \rangle$ in a Maxwell--Boltzmann distribution \cite{PhysRevD.96.123008}. In this procedure, we have considered an axion width equal to $\sim$1.5 times the full width half maximum of $S_a (\nu)$. The misalignment is defined over a range given by

\begin{equation}
           -z K_r \Delta \nu_b \le \delta \nu_r \le (1-z) K_r \Delta \nu_b \,
\end{equation}

for $0 \le z \le 1$. The value of $z$ must be set such that the sum $\sum_q L_q / K_g$ yields a larger value than the one obtained if the sum limits were shifted by 1 up or down for all $\delta \nu_r$. The weights $\bar{L}_q$ are obtained by averaging out $L_q (\delta \nu_r)$ over the range of $\delta \nu_r$. The grand spectrum is now calculated as

\begin{equation}
\label{eq:requisito1}
           D_\ell^g = \sum_q D_{\ell + q - 1}^r \bar{L}_q\,,
\end{equation}

and

\begin{equation}
\label{eq:requisito2}
           R_\ell^g = \sqrt{\sum_q \left ( R_{\ell + q - 1}^r \bar{L}_q \right )^2} \,.
\end{equation}

The normalised spectrum is obtained as $\delta_\ell^g / \sigma_\ell^g = D_\ell^g / R_\ell^g$. In order to account for filter-induced correlations among bins, the standard deviation, $\xi$, of the normalised grand spectrum is calculated, from which a corrected grand spectrum standard deviation $\tilde{\sigma}_\ell^g = {\sigma}_\ell^g / \xi$ is obtained. The output of this method, which is the normalised and corrected grand spectrum, is calculated as $\delta_\ell^g / \tilde{\sigma}_\ell^g$. An exclusion sector determined by the bins exceeding a threshold $\Theta$ for an arbitrary confidence level (CL) can be established, $\Theta = \mathrm{SNR} - \Phi^{-1}(\mathrm{CL})$, $\Phi$ being the cumulative distribution function of the standard normal distribution.

\begin{figure}[h]\centering
\includegraphics[width=0.45\textwidth]{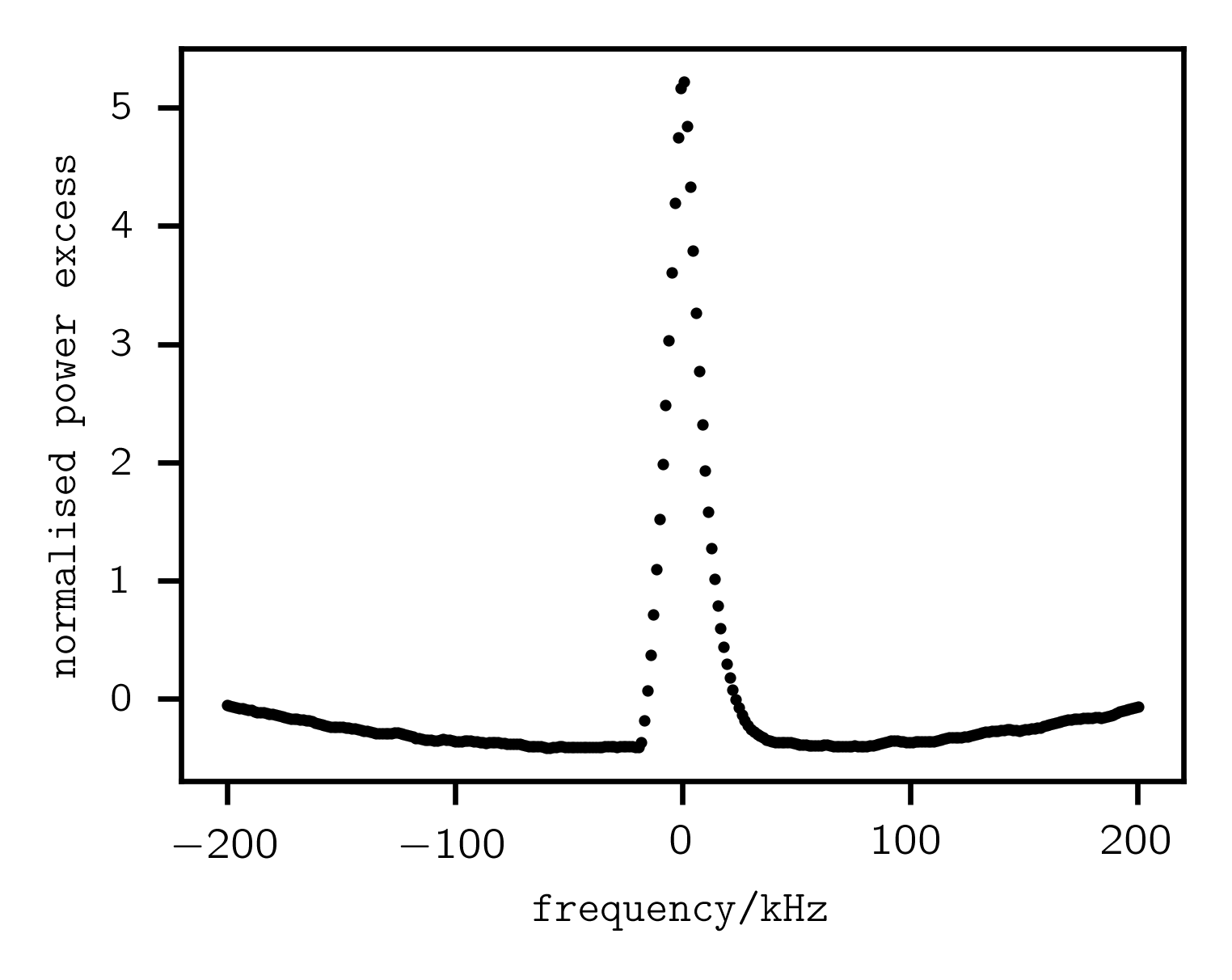}
\caption{Sensitivity calculation. The average value of the normalised corrected grand spectrum, ${\delta}_\ell^g / \tilde{\sigma}_\ell^g$, in a window of 150 bins to each side of the frequency where a synthetic axion signal has been injected is represented. The peak value (5.217) is to be interpreted as the SNR of the axion-induced signal. The distributions of the values of all bins of the analysed window have a standard deviation close to 1. The ripple around the peak is an artifact of the filtering. The abscissa axis has been scaled by $K_r$ to account for the rebinning.}
\label{fig_S1}
\end{figure}

In Fig. \ref{fig_S1}, we report the result of a simulation after  calculating 10,000 iterations. In each iteration, raw bins have been generated as random numbers in each spectrum drawn from a Gaussian distribution with mean $\mu = k_{\mathrm{B}} T_\mathrm{{sys}} \Delta \nu_b$ and deviation $\sigma = k_{\mathrm{B}} T_\mathrm{{sys}} \Delta \nu_b / \sqrt{t \Delta \nu_b}$ \cite{doi:10.1063/1.1770483}, scaled by the power boost curve \cite{De_Miguel_2021}. An artificially generated discrete axion-induced signal has been inserted to one of the spectra as $S_{a, \,ij} \approx \Delta \nu_b S_a (\nu_{i j})$, which is successfully flagged. The resulting spectra are introduced as an input to the analysis procedure and the value of the output in a window of 150 bins at each side of the bin where the axion signal was inserted has been recorded. An instantaneous scanning bandwidth of 50 MHz has been considered, effective integration time is set to five days per subspectrum, and an array size of $2^{18}$ has been considered resulting in a bin width $\Delta \nu_b \simeq 190.7$ Hz. A total of $M = 2$ spectra with an overlap of 5 MHz have been considered around a central frequency of about 24.2 GHz, where the synthetic axion signal has been injected with $g_{a \gamma \gamma} \simeq 1.5 \times 10^{-14} \, \mathrm{GeV}^{-1}$, which corresponds to a DFSZ I axion of 100 $\upmu$eV mass. Since an axion width of approximately 90 bins is expected at this frequency, $K_r = 7$ and $K_g = 14$ have been considered, for which the value $z = 0.51$ has been numerically found to best fit the requirements in Eq. \ref{eq:requisito1} and Eq. \ref{eq:requisito2}. The weights, $\bar{L}_q$, have been calculated for $\nu_a \simeq 24.2$ GHz, but any frequency in the scan range would be suitable given that variations of $S_a (\nu)$ along a series of $M$ subspectra tend to be negligible.  

\section{On the quality factor and the one-eighth wavelength plate stack}
In this material we explore the quality factor ($Q$) tenable in a dielectric Fabry--Pérot axion haloscope in practice, for which we put together the experimental results in \cite{Hernandez-Cabrera:2023syh, Hernandez-Cabrera:2024SPIE}. The devices under test (DUTs) are two similar fixed-plate Fabry--Pérot resonators with N= 2, ..., 5 layers of dimensions 100x100$\pm$0.5 mm size and 1$\pm$0.03 mm thickness each. The ceramic used is 3 mol\% yttria stabilized tetragonal zirconia \cite{PhysRevB.65.075105, PhysRevB.64.134301} polycrystal (3Y-TZP) aquired to SITUS Technicals GmbH. The surface roughness of the zirconia wafers is $<$$0.1$ $\upmu$m. Dielectric losses of this material are of the order of  $\mathord{\mathrm{tan}}\,\delta\sim10^{-4}$ at $\pazocal{O}(10)$ GHz frequencies \cite{10.1063/1.353963}. The value of the dielectric constant of this substrate obtained in our laboratory following the method in \textcite{https://doi.org/10.1002/mop.20390} turns out to be unusually low with respect to the typical permittivity of tetragonal zirconia (t-ZrO2) measured for a range of different compositions—in particular, $\varepsilon_r\sim10-14$ (measured) versus $\varepsilon_r\sim30-46$ \cite{ LANAGAN1989437, Miranda1990, 10.1063/1.353963}. Since the peak of the Lorentzian spectral feature of the quality factor scales with the permittivity of the dielectric boundary \cite{renk2012basics}, this commensurated the observed $Q$ by a factor of a few, although it does allow us to perform a proof of concept, including the experimental test of the one-eighth wavelength principle. In this concern, two different plate holders were manufactured in aluminum with an error of about three dozen microns or less, according to metrology data. The plate distance is 6.04 mm for DUT 1 and 6.21 mm for DUT 2 in order to resonate at frequencies $\sim$1/2 GHz apart. For each plate distance,  6.04 mm or 6.21 mm, the spacing is, simultaneously, a fraction of the scanning wavelength of $\sim$$\lambda/8$ for a one-eighth wavelength stack, to be tested in low frequency measurements, and $\sim$$\lambda/2$ for a half-wavelength configuration. 

\begin{figure}[H]\centering
\includegraphics[width=0.4\textwidth]{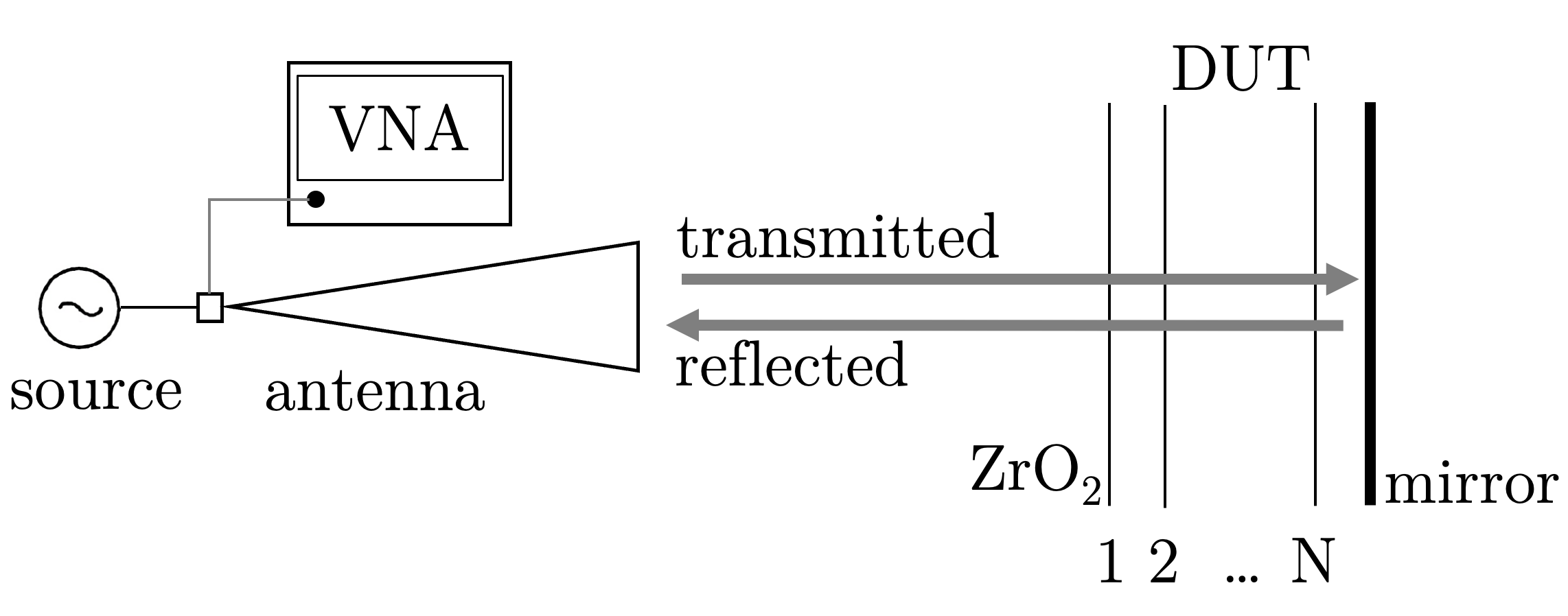}
\caption{Schematic setup used for the measurement of the quality factor of DUT1 and DUT2 at low and high frequency.}
\label{fig_SII1}
\end{figure}

\begin{figure}[H]\centering
\includegraphics[width=0.45\textwidth]{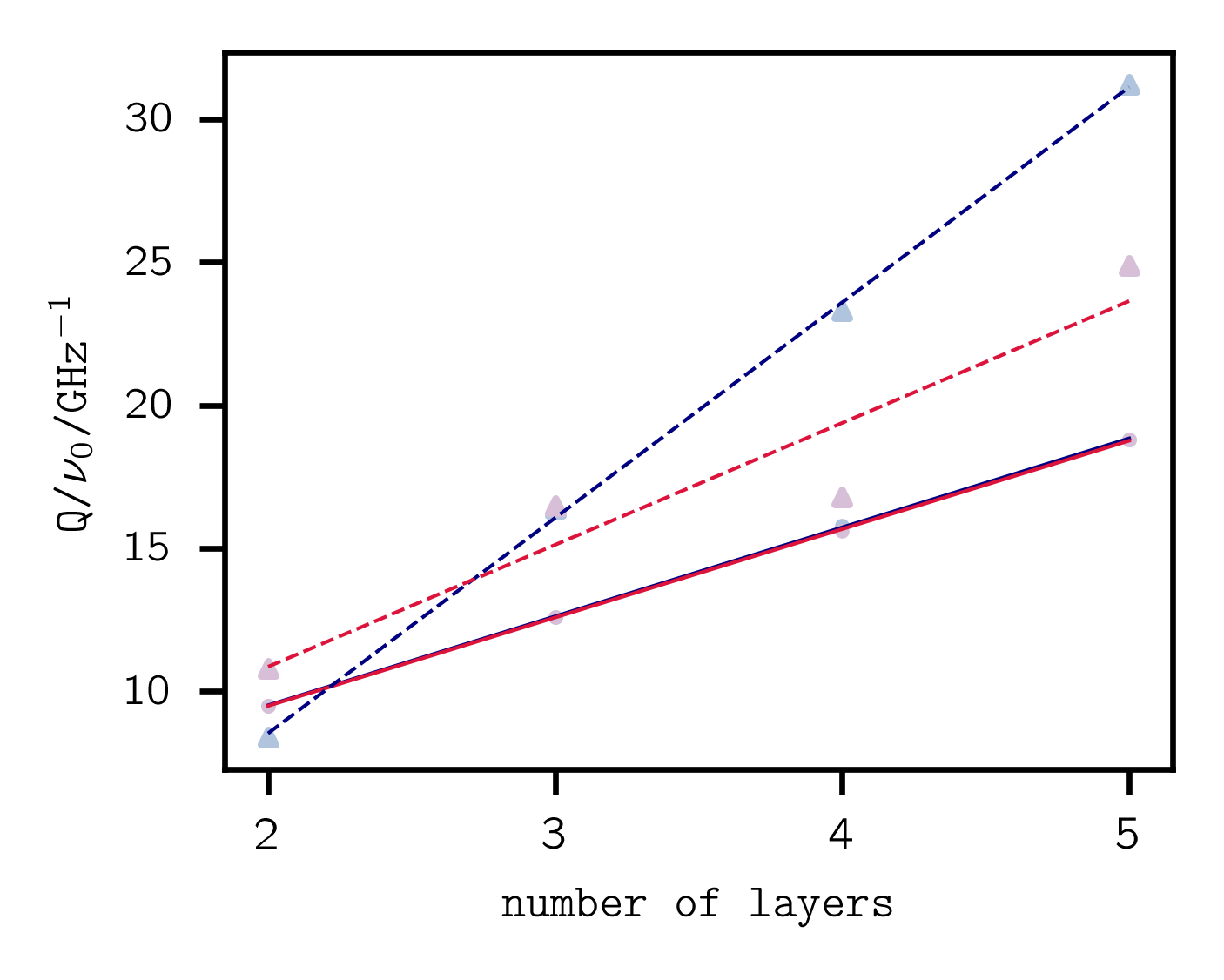}
\caption{Experimental measurements of the quality factor for a stack of N plates of 3 mol\% yttria stabilized tetragonal zirconia of 100x100x1 mm each. Two twin fixed-plate resonators were fabricated and tested with a frequency shift between them. The results for DUT 1 are shown in blue and those of DUT 2 in red. The circular points were taken at low frequency ($\sim$7 GHz) and the triangular points are at high frequency ($\sim$33 GHz) At high and low frequency, where diffraction can play a role since the plate size is of the order of the wavelength, different equipment is used. We normalize the observed quality factor to the resonant frequency to evidence the one-eight wavelength concept. The lines are the least squares fits. }
\label{fig_SII2}
\end{figure}

Both DUTs were tested at low and high frequency at room temperature. The alignment of the optical setup has an accuracy of $\lesssim$1$^{\circ}$. A vector network analyzer (VNA) is used to acquire the scattering parameters. Free-space measurement between two antennas are taken as a reference. Time domain data gating allows spurious reflections to be filtered out, surfacing the performance of each Fabry--Pérot resonator. When a flat copper mirror is inserted after the last plate maintaining the same spacing with it to maintain phase coherence, the reflection coefficient is measured and the group delay time of the photons through a DUT is estimated.  The quality factor is computed as $Q=-\omega d\phi/d\omega$, $\phi$ being the phase \cite{renk2012basics, Ismail:16}. A schema of the minimum assembly is shown in Fig. \ref{fig_SII1}.
The experiment results are shown in Fig. \ref{fig_SII2} \cite{Hernandez-Cabrera:2023syh, Hernandez-Cabrera:2024SPIE}. Inside each DUT, the incident wave and the reflected wave interfere coherently at both one-eighth and one-half of the scanning wavelength simultaneously, unveiling the power spectral density of the autocorrelation function at two different resonant frequencies that are approximately four wavelengths apart. In particular, the resonant feature peaks at about 7 and 33 GHz—in both DUTs, with the aforementioned relative shift. This is consistent with a periodic resonance over $\sim$$4\nu_0$ intervals once deviations with respect to an ideal plate thickness of $\sim$$\lambda/(8\sqrt{\varepsilon_r})$ or $\sim$$\lambda/(2\sqrt{\varepsilon_r})$ caused by the fixed plate thickness of 1 mm are considered.

From the inference that the quality factor scales linearly with the number of plates in series up to a relatively large number of layers \cite{Millar:2016cjp}, we estimate a maximum quality factor of $Q\gtrsim 10^4$ for a Fabry--Pérot axion haloscope with N$\sim$40--50 with this particular 3Y-TZP composition. Nevertheless, different t-ZrO$_2$ wafers with a different structure and composition provide a higher dielectric constant, of the order of $\varepsilon_r\gtrsim40$ \cite{10.1063/1.353963}. The quality factor for these substrates would scale up to $Q\gtrsim50,000$ at $\sim$60 GHz for a stack of $\sim$50 layers, even including here dielectric losses and a decrease of the dielectric constant observed at cryogenic temperatures \cite{Miranda1990}. Therefore, we are preparing to carry out further tests on a series of t-ZrO$_2$ ceramic substrates with better dielectric properties procured from various sources. This will allow us to determine the specific zirconia composition to be incorporated into DALI. 
 
\end{document}